**Diffusion mechanism of exciplex.**

**II. Energy transfer mechanism**

Hwang-Beom Kim,[1] and Jang-Joo Kim[1,2,*]

[1]*Department of Materials Science and Engineering and the Center for Organic Light Emitting Diodes, Seoul National University, Seoul 151-742, South Korea.*

[2]*Research Institute of Advanced Materials (RIAM), Seoul National University, Seoul 151-744, South Korea.*

**Abstract**

Excited-state charge-transfer complexes (exciplexes) have been actively exploited in organic optoelectronic devices to improve performance; however, diffusion of exciplexes has not been actively studied despite its influence on performance due to the lack of apparent charge-transfer absorption. In the preceding paper, we studied the energy transfer (ET) from exciplexes to exciplex-forming pairs in relation to the charge-transfer absorption of the exciplex state, resulting in the exciplex diffusion. In this paper, we report that the ET takes place dominantly via the Dexter-type exchange mechanism, from the exponential decrease of the ET rate constant with separation between exciplexes.

**1. Introduction**

There are a couple of papers describing the diffusion of exciplex or polaron pairs. C. Deibel et al. suggested that the diffusion of polaron pairs might account for the efficiency loss in organic photovoltaics (OPVs) by applying Monte Carlo simulations [1]. M. A. Baldo and coworkers have reported that charge transfer (CT) exciplex states can be transported via an "inchworm"-type mechanism [2]. It was discussed that Förster-type resonance energy transfer (FRET) for diffusion of

CT exciplex states would not be effective because of the negligibly low optical absorption of CT exciplex states, and that Dexter-type exchange energy transfer (DET) would also not be effective because CT exciplex states are weakly bound.

There are two requirements for DET to take place by the Fermi golden rule. One is the electronic exchange integral between energy donor states and energy acceptor states. The electronic exchange integral drops exponentially with the separation between them, and DET was thought to occur when the energy donor and acceptor are in physical contact [3]. However, the physical contact is not always the necessary condition for efficient long-range DET to occur, and the long-range DETs without the physical contact have been report, known as the superexchange energy transfer (ET) [4–6]. The large coupling between π-conjugated molecules could expand the exchange integral between them to large separation when they (donor and acceptor) are π-conjugated molecules in the π-conjugated medium. It indicates that the physical contact between the energy donor and acceptor is not necessary for DET. The other is the spectral overlap between absorption of energy acceptors and emission of energy donors which is the experimental view of the density of the isoenergetic energy-accepting states with energy donor states. Thus, ET mechanisms (DET as well as FRET) have not been considered for exciplex-forming systems because of the lack of apparent CT absorption in standard steady-state absorption measurements in exciplex-forming systems. However, there are experimental results on the existence of sub-bandgap CT absorption in exciplex-forming systems with very low intensities employing elaborate measurements [7–9]. It should be noted that the DET rate constant does not depend on the magnitude of the transition dipole moment of the energy acceptor, so that the considerable DET rate constant is possible even with so low extinction coefficient for the CT absorption unlike FRET [10].

In the first paper (Paper I), we discussed that the energy transfer from exciplexes to exciplex-forming pairs takes place by the transient analysis. Here, we show that DET is the dominant mechanism for exciplex diffusion. The exponential decrease of the ET rate constant with separation and the existence of the sub-bandgap CT absorption in exciplex-forming films indicate that the dominant ET mechanism from exciplexes to exciplex-forming pairs is the Dexter exchange mechanism even though the CT absorption is very weak because DET is independent of the oscillator strength of the energy acceptor [7–9, 11–13].

## 2. Results

To investigate the relationship between the ET rate constant and the distance between the energy donor and acceptor, we conducted concentration-dependent exciplex-quenching experiments with various molar ratios. Total 16 films (doped films) were fabricated: eight TCTA:PO-T2T films with molar ratios of $x$: 100 - $x$ and eight TCTA:m-MTDATA:PO-T2T films with molar ratios of $x$: $x$: $100 - 2x$, respectively, where $x$ = 3.0, 3.5, 4.0, 4.5, 5.0, 5.5, 6.0, and 6.5. TCTA, m-MTDATA, and PO-T2T represent 4,4′,4″-tri(*N*-carbazolyl)triphenylaminem 4,4′,4″-tris (3-methyl-phenylphenylamino) triphenylamine, and 4,4′-bis(3-methylcarbazol-9-yl)-2,2′-biphenyl, respectively. The doping concentrations of the films were kept low to maintain identical local environments for the dopants and ensure that the photo-physics of the TCTA:PO-T2T exciplex was not influenced by the addition of m-MTDATA molecules. The grazing incident small angle X-ray scattering measurement indicates that TCTA and m-MTDATA molecules do not form aggregates in the films, as shown in the Supplemental Material (Fig. S1) [14, 15]. The transient photoluminescence (PL) intensities of the films are shown in Fig. 1. The transient PL profiles were fitted by a bi-exponential decay model [16]. The ET rate constant from the TCTA:PO-T2T exciplexes to the m-MTDATA:PO-T2T exciplex-forming pairs, $k_{ET}$, was calculated as the difference between the prompt decay rate constants of the TCTA:PO-T2T exciplexes in the TCTA:PO-T2T and TCTA:m-MTDATA:PO-T2T films with the same TCTA concentrations like Paper I [16], which is plotted in Fig. 2(a).

## 3. Discussion

DET can be considered as a mechanism for the ET because the extinction coefficient of the CT absorption for the exciplexes is very small in the order of $10^{-3}$. The discuss on DET was carried out based on the relationship between the ET rate constant and separation. DET between the nearest energy donor-acceptor pairs was considered in the analysis. DET between more separated pairs was ignored because the DET rate constant decreases exponentially with the separation. TCTA and m-MTDATA molecules doped into bulk of PO-T2T molecules at low concentrations were assumed to be uniformly dispersed. The DET equation can then be written as follows, as explained in the Supplemental Material [17, 18].

$$\log k_{DET} = \log(NKJ) - \frac{\beta}{2.303}\left(\frac{3}{\pi N_A}\right)^{1/3} C_{\text{m-MTDATA}}^{-1/3} \tag{1}$$

Here, $k_{DET}$ is the DET rate constant, $N$ the number of nearest energy-accepting species, $K$ the experimental factor related to the specific orbital interactions, $J$ the spectral overlap integral normalized for the molar extinction coefficient of the energy-accepting species, $\beta$ the attenuation factor arising from the electronic exchange integral that is considered to drop exponentially in the tail of molecular orbitals [3, 4], $N_A$ Avogadro's number, and $C_{\text{m-MTDATA}}$ the molar concentration of m-MTDATA in the TCTA:m-MTDATA:PO-T2T film.

The linear relationship of the semilogarithmic plot in Fig. 2(a) indicates that the ET rate constant decreases exponentially with the distance between the TCTA:PO-T2T exciplex and the m-MTDATA:PO-T2T exciplex-forming pair in the mixed films, demonstrating that ET takes place by DET. The attenuation factor arising from the electronic exchange integral was 1.1 nm$^{-1}$, and the pre-exponential constant was $NKJ = 2.9\times10^8$ s$^{-1}$. S. R. Forrest et al. reported that the diffusion of the triplet excited states of phosphors takes place via Dexter-type exchange mechanism [5]. When tris[2-phenylpyridinato-C$^2$,$N$]iridium(III) (Ir(ppy)$_3$), bis[2-(2-pyridinyl-$N$)phenyl-C](acetylacetonato)iridium(III) (Ir(ppy)$_2$acac), and platinum octaethylporphyrin (PtOEP) are each doped in 4,4'-bis(carbazol-9-yl)biphenyl (CBP) host film, excited dopant molecules transfer their energies to neighbor dopant molecules via Dexter-type exchange mechanism, and they have the attenuation factors of 1.25, 1.2, and 1.5 nm$^{-1}$, respectively, and the pre-exponential constants of 2.3×10$^8$ s$^{-1}$, 3.0×10$^8$ s$^{-1}$, and 6.4×10$^6$ s$^{-1}$, respectively, which are the similar values with our results [5]. It indicates that the ET from exciplexes to exciplex-forming pairs also takes place by Dexter exchange mechanism in organic films. The DET process from exciplexes to exciplex-forming pairs can be schematically represented by Fig. 2(b) in the view of the molecular-orbital diagram. The state energy diagram represents the long-range DET with small attenuation factors in which $v_1$, $v$, and $v_2$ are the electronic exchange integral between the exciplex and medium molecule, that between the medium molecules, and that between the medium molecule and exciplex-forming pair, respectively, as shown in Fig. 2(c) [6, 19–21]. As a result of the first-order perturbation theory, $V_{total}$ is the superexchange electronic exchange integral which is the indirect electronic exchange integral via intermediate states (medium states) [5, 20–22]. The more detailed analysis of the ET described in the Supplemental Material. The DET process would be dominant in polymer systems as well, considering that absorption coefficients for the CT state in the polymer:[6,6]-phenyl C$_{61}$ butyric acid methyl ester (PCBM) mixed films exploited in OPVs are similar with that in the m-MTDATA:PO-T2T film as shown in Fig. S2 [7]. More details are described in the Supplemental Material [23].

The other various mechanisms also can be considered for the ET process from exciplexes to exciplex-forming pairs. However, ET rate constants via the other mechanisms for the ET process other than the exchange mechanism were analyzed to be negligible in our investigated exciplex-forming system.

First, FRET was considered. The FRET rate constant ($k_{FRET}$) can be expressed by the following equation [24].

$$k_{FRET} = \frac{9000 \ln 10}{128 \pi^5 N_A} \frac{N_R \kappa^2 k_r}{n^4} \frac{\int I_D(\lambda) \varepsilon_A(\lambda) \lambda^4 d\lambda}{\int I_D(\lambda) d\lambda} \frac{1}{R^6} \quad (2)$$

where $k_{FRET}$ is the FRET rate constant, $N_R$ is the number of energy acceptors with separation of $R$

from an energy donor, $\kappa^2$ is the dipole orientation factor, $k_r$ is the radiative rate constant of the energy-donating states, $n$ is the refractive index of the medium, $\lambda$ is the wavelength (in units of cm), $I_D$ is the emission spectrum of the energy-donating states, $\varepsilon_A$ is the molar extinction coefficient of energy-accepting states (in units of L cm$^{-1}$ mol$^{-1}$), and $R$ is the distance between the energy donors and acceptors (in units of cm). Based on the assumption that TCTA and m-MTDATA molecules doped into bulk of PO-T2T molecules at low concentrations are uniformly dispersed in the Wigner-Seitz approximation [18], the number of the nearest energy-accepting species is 6 in TCTA:m-MTDATA:PO-T2T doped films. We only consider the FRET between the nearest energy donor-acceptor pairs because the FRET between the second nearest energy donor-acceptor pairs would be more than 10 times smaller than that. The dipole orientation factor was assumed to be 2/3 for randomly oriented energy donors and energy acceptors. The refractive index of the medium of TCTA:m-MTDATA:PO-T2T doped films was determined to be 1.73 which is the refractive index of PO-T2T neat films at the wavelength of 500 nm. The nearest distance between TCTA:PO-T2T exciplexes and m-MTDATA:PO-T2T exciplex-forming pairs could be calculated by Eq. (S1). The emission spectrum of the TCTA:PO-T2T singlet exciplex is depicted in Fig. 2(b) of Paper I. The molar extinction coefficient of the m-MTDATA:PO-T2T singlet exciplex (in units of L cm$^{-1}$ mol$^{-1}$) could be obtained by Eq. (3).

$$\varepsilon_A = \frac{4\pi\kappa}{\lambda} C_{\text{m-MTDATA}}^{-1} = \frac{4\pi\kappa}{\lambda} \left( \frac{1000\rho_{\text{film}}}{M_w} \right)^{-1} \tag{3}$$

where $\kappa$ is the extinction coefficient of m-MTDATA:PO-T2T singlet exciplexes as shown in Fig. 2(b) of Paper I [25], $\rho_{\text{film}}$ is the density of the film (in units of g cm$^{-3}$), and $M_W$ is the sum of molecular weights of m-MTDATA and PO-T2T (in units of g mol$^{-1}$) for the m-MTDATA:PO-T2T (m.r 1:1) film. All films were assumed to have a density of 1 g cm$^{-3}$. The photoluminescence quantum yield (PLQY) of TCTA:PO-T2T exciplexes was measured to be 35%. The PLQY for exciplex emission with TADF characteristics is expressed by the following equation [16],

$$\Phi_{PL} = \frac{k_r}{k_{prompt} - k_{isc}\Phi_{risc}} \tag{4}$$

where $\Phi_{PL}$ is the PLQY of exciplexes, $k_r$, $k_{prompt}$, and $k_{isc}$ are the fluorescence rate constant, prompt decay rate constant, and intersystem crossing rate constant from singlet exciplex states, respectively, and $\Phi_{risc}$ is the reverse intersystem crossing quantum yield. The maximum $k_r$ of the TCTA:PO-T2T singlet exciplex is $1.0 \times 10^7$ s$^{-1}$ based on the prompt decay rate constants of the TCTA:PO-T2T films in Table S1 when $\Phi_{risc} = 0$, which cannot be the real value because the delayed fluorescence is observed in the TCTA:PO-T2T films. We set the radiative rate constants of the TCTA:PO-T2T singlet exciplex state as $1.3 \times 10^6$, $2.5 \times 10^6$, $5.0 \times 10^6$, and $1.0 \times 10^7$ s$^{-1}$, and plotted the Eq. (2) as dashed lines in Fig. 3(a). The FRET rate constants are several times smaller than the DET rate constants. Not only that, the concentration dependence of the FRET rate constants does not follow the experimental data [$\propto R^{-6}$ for FRET vs. exp $(-\beta R)$ for DET]. The much smaller FRET rate constant of the exciplex can be understood from the very small extinction coefficient for CT absorption. These results reveal that FRET is not the dominant mechanism for the exciplex diffusion.

Second, exciplex dissociation, followed by trap assisted recombination, can be considered as a mechanism of exciplex diffusion. ET via trap-assisted recombination of polarons formed by the dissociation of exciplexes could occur. TCTA:PO-T2T exciplexes dissociate into polarons, and then trap-assisted recombination occurs to form m-MTDATA:PO-T2T exciplexes, as depicted in Eq. (5).

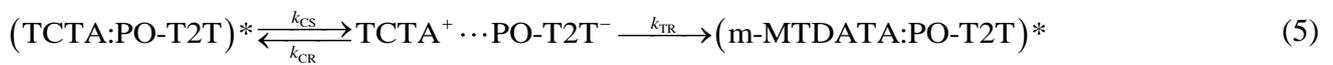

$$(\text{TCTA:PO-T2T})^* \underset{k_{CR}}{\overset{k_{CS}}{\rightleftarrows}} \text{TCTA}^+ \cdots \text{PO-T2T}^- \xrightarrow{k_{TR}} (\text{m-MTDATA:PO-T2T})^* \tag{5}$$

When a steady-state approximation for the concentrations of intermediates is applied which is described in Eq. (6), the quenching rate constant of TCTA:PO-T2T exciplexes in the above reaction could be expressed by Eq. (7).

$$k_{CS}[(\text{TCTA:PO-T2T})^*] - k_{CR}[\text{TCTA}^+ \cdots \text{PO-T2T}^-] - k_{TR}[\text{TCTA}^+ \cdots \text{PO-T2T}^-] = 0 \tag{6}$$

$$k_{ET}\left[(TCTA:PO\text{-}T2T)^*\right] = k_{CS}\left[(TCTA:PO\text{-}T2T)^*\right] - k_{CR}\left[TCTA^+ \cdots PO\text{-}T2T^-\right]$$
$$= k_{CS}\left(1 - \frac{1}{k_{TR}/k_{CR}+1}\right)\left[(TCTA:PO\text{-}T2T)^*\right] \quad (7)$$

$k_{TR}$ could be expressed by the following equation [26].

$$k_{TR} = \frac{q\mu_n n_{trap}}{\varepsilon\varepsilon_0} = \frac{q\mu_n N_A C_{m\text{-}MTDATA}}{\varepsilon\varepsilon_0} \quad (8)$$

where $q$ is the elementary charge, $\mu_n$ is the electron mobility in the film (in units of cm$^2$ V$^{-1}$ s$^{-1}$), $n_{trap}$ is the density of the traps for holes corresponding to the density of m-MTDATA in the TCTA:m-MTDATA:PO-T2T mixed film, $C_{m\text{-}MTDATA}$ is the molar concentration of m-MTDATA in the films (in units of mol cm$^{-3}$), $\varepsilon$ is the dielectric constant of the film, and $\varepsilon_0$ is the vacuum permittivity. We considered that m-MTDATA molecules behave as hole traps when the trap-assisted recombination engages in the energy migration from the TCTA:PO-T2T exciplexes to the m-MTDATA:PO-T2T exciplex-forming pairs in the TCTA:m-MTDATA:PO-T2T films, not only because m-MTDATA has higher HOMO levels than TCTA and PO-T2T, but also logically the positive polaron originated from TCTA must be trapped in m-MTDATA for the ET to take place from the TCTA:PO-T2T exciplexes to the m-MTDATA:PO-T2T exciplex-forming pairs by the trap assisted recombination. Thus, we used the molecular density of m-MTDATA in the films as the trap density. For our specific system of TCTA:m-MTDATA:PO-T2T system with $\varepsilon = 4$ and $\mu_n = 1.2 \times 10^{-5}$ cm$^2$/V·s [27], Eq. (7) becomes,

$$k_{ET} = k_{CS}\left(1 - \frac{1}{3.3\times10^{12}/k_{CR} \times C_{m\text{-}MTDATA}+1}\right) \quad (9)$$

The value of $k_{CS}$ and $k_{CR}$ are not known for the TCTA:m-MTDATA:PO-T2T film. Instead of determining the values by experiment, we used approximate values taken from a similar exciplex system in literature, where an 1:1 mixed film of m-MTDATA:3TPYMB was used as an exciplex-forming mixed film to report $k_{CS}/k_{CR} = 1.36\times10^{-3}$ and the binding energy of the m-MTDATA:3TPYMB exciplex of 0.17 eV [28]. If we assume the exciplex binding energies in small-molecular π-conjugated organic films are similar, then $k_{CS}/k_{CR} = 1.36\times10^{-3}$ must be similar in the both systems. Using the relationship, the $k_{ET}$ values are calculated for different $k_{CR}$ values of $5\times10^7$ s$^{-1}$, $1\times10^8$ s$^{-1}$, $5.3\times10^8$ s$^{-1}$ [26], $1\times10^9$ s$^{-1}$, and $5\times10^9$ s$^{-1}$, and are compared with the experimental data and DET fitting in Fig. 3(b). The ET rate constants via trap-assisted recombination of polaron pairs generated from the dissociation of exciplexes are almost two orders of magnitude smaller than the experimental ET rate constants which are fitted by the Dexter-type exchange mechanism. Therefore, exciplex diffusion via trap-assisted recombination must be negligible compared to the DET.

Third, we consider the inchworm mechanism of exciplex diffusion reported in ref. 2 based on the magneto-PL observed from exciplex and Monte Carlo simulation, where an electron and a hole from an exciplex hop to neighboring molecules incoherently, and recombine geminately. It may be a feasible mechanism. Further study is required to clarify whether the inchworm mechanism can be applicable to the systems in this study. In the reference, the magnetic effect was interpreted to originate from the stretching of an electron-hole pair, followed by the hyperfine interaction (HFI) of the elongated electron-hole pair. However, the origin of the magnetic effect for exciplexes is controversial at this moment. Δg mechanism and HFI are under consideration as the dominant origin of the magnetic effect for exciplexes [29, 30], In addition, the ET rate constants from exciplexes to exciplex-forming pairs are too large to be explained by the inchworm mechanism. The inchworm mechanism has three steps; dissociation of exciplexes into electrons and holes, incoherent hopping of the electrons and holes in the bound state and geminate recombination of the electrons and holes. The processes are sequential so that the slowest step must be the rate determining step. M. A. Baldo and coworkers have reported the rate constant for the exciplex dissociation into the polaron pair as $7.2\times10^5$ s$^{-1}$ and the geminate recombination of the polaron pairs as $5.3\times10^8$ s$^{-1}$, respectively [28]. Therefore, the ET rate constant via the inchworm mechanism must be limited to $7.2\times10^5$ s$^{-1}$ no matter how large the rate constant for charge hopping after exciplex dissociation is. Even though the exciplex dissociation rate constant can be different for different exciplexes, this value is too small to explain our reported ET values with the maximum value of $1.3\times10^7$ s$^{-1}$ [9, 28]. Furthermore, it is difficult for the mechanism to describe the exponential decrease with the separation when the dissociation of exciplexes into electrons and holes is the rate determining step. Therefore, the inchworm mechanism would not be a dominant mechanism in the exciplex-forming systems in this study. More details are described in the Supplemental Material [31, 32].

**4. Conclusion**

In conclusion, the ET from exciplexes to exciplex-forming pairs can be explained by DET according to a concentration-dependent quenching experiment. We concluded that DET could be the dominant mechanism for exciplex diffusion in our investigated exciplex-forming systems after investigating other possibilities. The diffusion mechanism must be widely applied for most of optoelectronic materials in organic light-emitting diodes (OLEDs) and OPVs. As the importance of exciplexes or CT states in organic photonic devices such as OLEDs and OPVs has increased in recent years, the ET process discussed in this study will play an important role in understanding device characteristics and will contribute to the design of molecules and devices with improved performance.

**Acknowledgements**

S. W. Lee and H. J. Kim carried out the GISAXS measurement. This work was supported by the Mid-career Researcher Program (2018R1A2A1A05018455) through a National Research Foundation (NRF) grant funded by the Ministry of Science and ICT (MSIT).

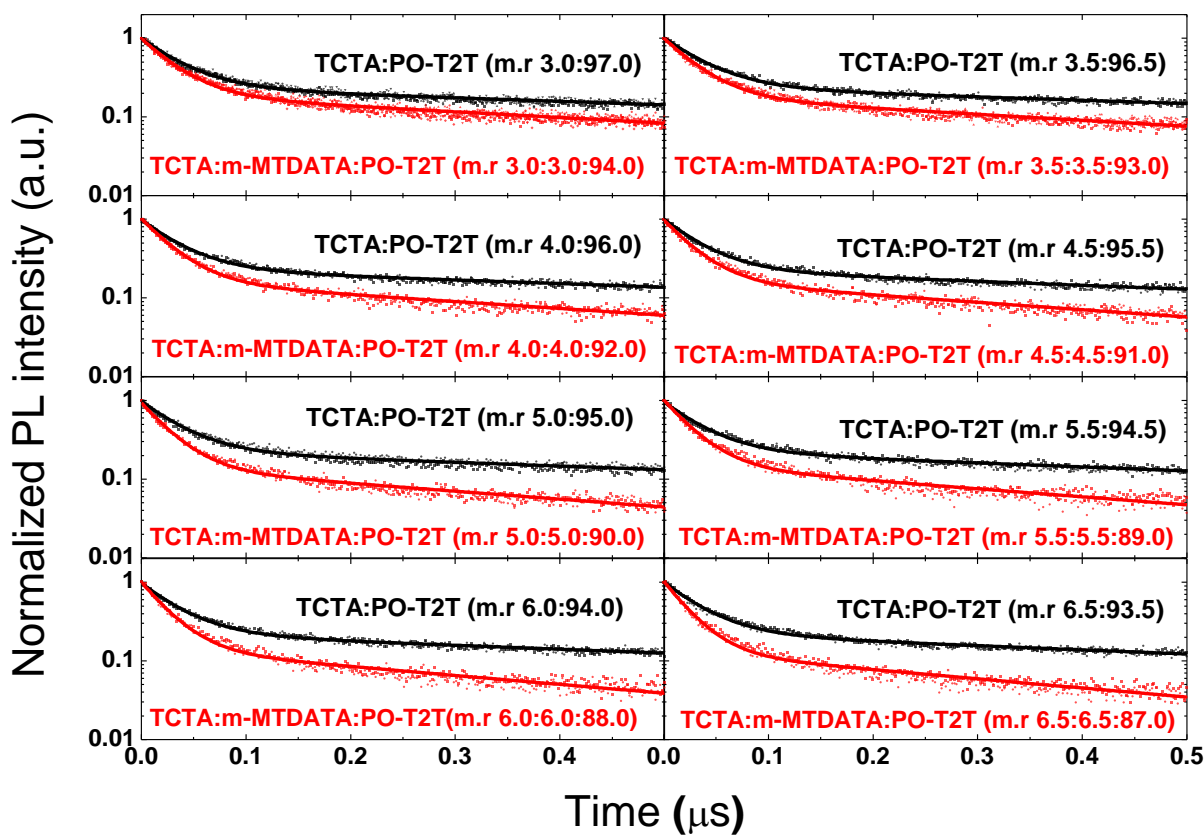

FIG. 1. Normalized transient PL intensities in the wavelength range 510 to 530 nm, where the PL of the TCTA:PO-T2T exciplex is dominant in TCTA:PO-T2T films and TCTA:m-MTDATA:PO-T2T films with various molar ratios (points), and fit lines by a two-exponential decay model (line) using the parameters in Table S1.

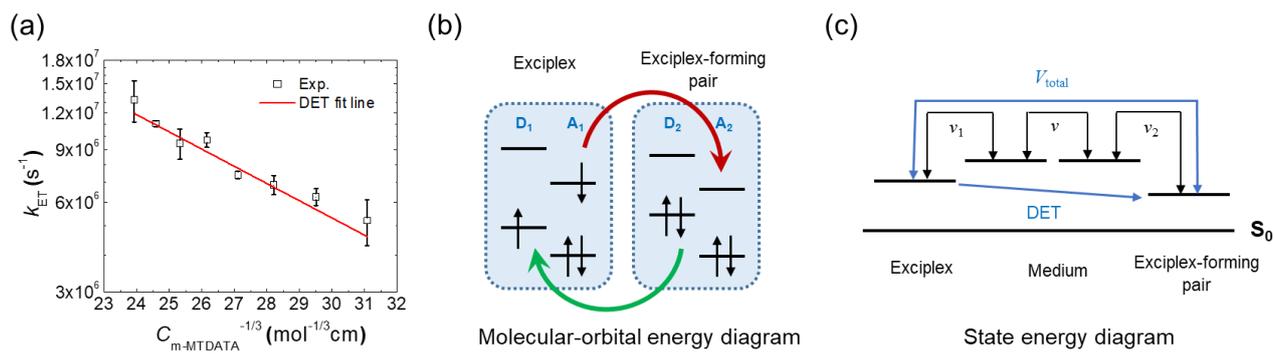

FIG. 2. (a) A plot of experimental values (square) of the logarithm scale of $k_{ET}$ vs. $C_{\text{m-MTDATA}}^{-1/3}$, and a Dexter-type exchange energy transfer (DET) fit line (solid line). Schematic (b) molecular-orbital and (c) state energy diagrams of DET from exciplexes to exciplex-forming pairs in π-conjugated molecular films.

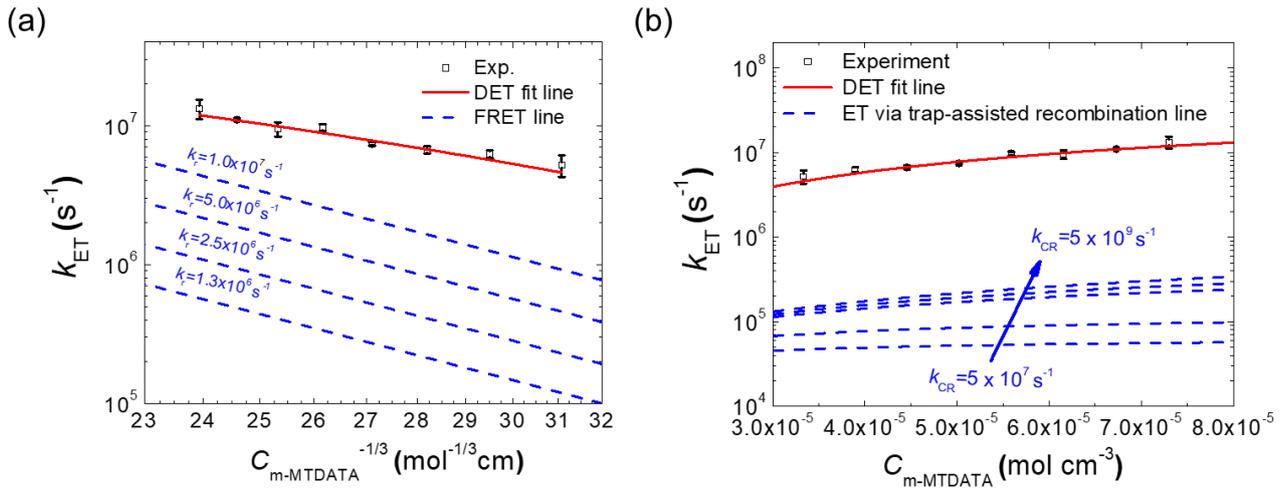

FIG. 3. (a) A plot of experimental values (empty square) of the logarithm scale of $k_{ET}$ vs. the logarithm scale of $C_{\text{m-MTDATA}}^{-1/3}$, a Dexter-type exchange energy transfer (DET) fit line (solid line), and possible lines of Förster-type resonance energy transfer (FRET) rate constants (Eq. (2)) (dashed line). (b) A plot of experimental values (empty square) of logarithm scale of $k_{ET}$ vs. $C_{\text{m-MTDATA}}$, a Dexter ET fit line (solid line), and possible lines of rate constants for ET via trap-assisted recombination (Eq. (9) when $k_{CR}$ values are $5\times10^7$ s$^{-1}$, $1\times10^8$ s$^{-1}$, $5.3\times10^8$ s$^{-1}$ [26], $1\times10^9$ s$^{-1}$, and $5\times10^9$ s$^{-1}$) (dashed line).